\documentclass[a4paper,11pt]{article}
\pdfoutput=1 

\usepackage{jinstpub} 

\title{\boldmath New Eco-gas mixtures for the Extreme Energy Events MRPCs: results and plans}

\author[1,4,*]{S. Pisano,\note[*]{Corresponding author, email: silvia.pisano@lnf.infn.it}}
\author[1,2]{M.~Abbrescia,}
\author[1,3]{C.~Avanzini,}
\author[1,3]{L.~Baldini,}
\author[1,4]{R.~Baldini Ferroli,}
\author[1,3]{L G.~Batignani,}
\author[1,17]{M.~Battaglieri,}
\author[1,8]{S.~Boi,}
\author[1,5]{E.~Bossini,}
\author[1,6]{F.~Carnesecchi,}
\author[1,7]{A.~Chiavassa,}
\author[1,8]{C.~Cicalo,}
\author[1,6]{L.~Cifarelli,}
\author[1]{F.~Coccetti,}
\author[1,9]{E.~Coccia,}
\author[1,10]{A.~Corvaglia,}
\author[1,11]{D.~De~Gruttola,}
\author[1,11]{S.~De Pasquale,}
\author[1,4]{L.~Fabbri,}
\author[16]{ V.~Frolov,}
\author[1,7]{L.~Galante,}
\author[1,7]{P.~Galeotti,}
\author[1,6]{M.~Garbini,}
\author[1,17]{G.~Gemme,}
\author[1,7]{I.~Gnesi,}
\author[1]{ S.~Grazzi,}
\author[1,12]{C.~Gustavino,}
\author[1,6,15]{D.~Hatzifotiadou,}
\author[1,18]{P.~La~Rocca,}
\author[1,19]{G.~Mandaglio,}
\author[14]{O.~Maragoto Rodriguez,}
\author[13]{G.~Maron,}
\author[1,20]{M.~N.~Mazziotta,}
\author[1,4]{S.~Miozzi,}
\author[1,6]{R.~Nania,}
\author[1,6]{F.~Noferini,}
\author[1,21]{F.~Nozzoli,}
\author[1,6]{F.~Palmonari,}
\author[1,10]{M.~Panareo,}
\author[1,10]{M.~P.~Panetta,}
\author[1,5]{R.~Paoletti,}
\author[14]{W.~Park,}
\author[1,6]{C.~Pellegrino,}
\author[1,17]{L.~Perasso,}
\author[1,3]{F.~Pilo,}
\author[1,7]{G.~Piragino,}
\author[1,18]{F.~Riggi,}
\author[1]{G.~C.~Righini,}
\author[1,11]{C.~Ripoli,}
\author[1,2]{M.~Rizzi,}
\author[1,6]{G.~Sartorelli,}
\author[1,6]{E.~Scapparone,}
\author[1,22]{M.~Schioppa,}
\author[1,3]{A.~Scribano,}
\author[1,6]{M.~Selvi,}
\author[1,8]{S.~Serci,}
\author[1,17]{S.~Squarcia,}
\author[1,17]{M.~Taiuti,}
\author[1,3]{G.~Terreni,}
\author[1,23]{A.~Trifir\`{o},}
\author[1,23]{M.~Trimarchi,}
\author[13]{M.~C.~Vistoli,}
\author[1,12]{L.~Votano,}
\author[1,6,15]{M.~C.~S.~Williams,}
\author[1,14,15]{L.~Zheng,}
\author[1,6,15]{A.~Zichichi,}
\author[1,15]{R.~Zuyeuski}

\affiliation[1]{Museo Storico della Fisica e Centro Studi e Ricerche Enrico Fermi, Roma, Italy}
\affiliation[2]{INFN and Dipartimento Interateneo di Fisica, Universit\`{a}
        di Bari, Bari, Italy}
\affiliation[3]{INFN and Dipartimento di Fisica, Universit\`{a} di Pisa,
        Pisa, Italy}
\affiliation[4]{INFN, Laboratori Nazionali di Frascati, Frascati (RM),
        Italy}
\affiliation[4]{INFN, Laboratori Nazionali di Frascati, Frascati (RM),
        Italy}
\affiliation[5]{INFN Gruppo Collegato di Siena and Dipartimento di Fisica,
        Universit\`{a} di Siena, Siena, Italy}
\affiliation[6]{INFN and Dipartimento di Fisica e Astronomia,
        Universit\`{a} di Bologna, Bologna, Italy}
\affiliation[7]{INFN and Dipartimento di Fisica, Universit\`{a} di Torino,
        Torino, Italy}
\affiliation[8]{INFN and Dipartimento di Fisica, Universit\`{a} di
        Cagliari, Cagliari, Italy}
\affiliation[9]{INFN and Dipartimento di Fisica, Universit\`{a} di Roma Tor
        Vergata, Roma, Italy}
\affiliation[10]{INFN and Dipartimento di Matematica e Fisica,
        Universit\`{a} del Salento, Lecce, Italy}
\affiliation[11]{INFN and Dipartimento di Fisica, Universit\`{a} di
        Salerno, Salerno, Italy}
\affiliation[12]{INFN, Laboratori Nazionali del Gran Sasso, Assergi (AQ),
        Italy}
\affiliation[13]{INFN CNAF, Bologna, Italy}
\affiliation[14]{ICSC World Laboratory, Geneva, Switzerland}
\affiliation[15]{CERN, Geneva, Switzerland}
\affiliation[16]{JINR Joint Institute for Nuclear Research, Dubna, Russia}
\affiliation[17]{INFN and Dipartimento di Fisica, Universit\`{a} di Genova,
        Genova, Italy}
\affiliation[18]{INFN and Dipartimento di Fisica e Astronomia,
        Universit\`{a} di Catania, Catania, Italy}
\affiliation[19]{INFN Sezione di Catania and Dipartimento di Scienze
        Chimiche, Biologiche, Farmaceutiche e Ambientali, Universit\`{a} di
        Messina, Messina, Italy}
\affiliation[20]{INFN Sezione di Bari, Bari, Italy}
\affiliation[21]{Trento Institute for Fundamental Physics and Applications,Trento, Italy}
\affiliation[22]{INFN and Dipartimento di Fisica, Universit\`{a} della
        Calabria, Cosenza, Italy}
\affiliation[23]{INFN Sezione di Catania and Dipartimento di Scienze Matematiche e Informatiche, Scienze Fisicha e Scienze della Terra, Universit\'a di Messina, Messina, Italy}
\emailAdd{collaborazioneEEE@centrofermi.it}
\abstract{The Extreme Energy Events observatory is an extended muon telescope array, covering more than 10 degrees both in latitude and longitude. Its 59 muon telescopes are equipped with tracking detectors based on Multigap Resistive Plate Chamber technology with time resolution of the order of a few hundred picoseconds. The recent restrictions on greenhouse gases demand studies for new gas mixtures in compliance with the relative requirements. Tetrafluoropropene is one of the candidates for tetrafluoroethane substitution, since it is characterized by a Global Warming Potential around 300 times lower than the gas mixtures used up to now. Several mixtures have been tested, measuring efficiency curves, charge distributions, streamer fractions and time resolutions. Results are presented for the whole set of mixtures and operating conditions,
focusing on identifying a mixture with good performance at the low rates typical of an EEE telescope.
}

\arxivnumber{XXXX.XXXX} 
\collaboration{The EEE collaboration}
\proceeding{The XIV$^{\text{th}}$ Workshop on Resistive Plate Chambers and related detectors\\
Feb. 19-23/2018 \\
Puerto Vallarta, Jalisco State, MEXICO}
\begin{document}
\maketitle
\flushbottom
\section{Introduction}
\label{sec::intro}
The Extreme Energy Events (EEE) \cite{eee_project} experiment is a strategic project led by the italian Museo Storico della Fisica e Centro Studi e Ricerche "Enrico Fermi" (Centro Fermi \cite{CFsite}), carried on in collaboration with INFN, CERN and MIUR, aiming at the detection and analysis of highly energetic cosmic rays. The experiment consists of a network of 59 muon telescopes, based on the Multigap Resistive Plate Chamber (MRPC in the following) technology, and already produced relevant observations in the field \citep{coinc1,coinc2,forbushDec2,forbushDec3,anisotropies,upward}. Due to its wide coverage over the Italian territory (more than 10 degrees both in latitude and in longitude, covering an area larger than 3$\times$10$^5$ km$^2$), the EEE network is the largest MRPC-based system for
cosmic rays detection. Details on the detector performances can be found in \cite{proc_ma,proc_ddg,eee_perf_paper}.
The current MRPCs are six gas gaps detectors, 300 $\mu$m each. The chambers are filled with a mixture made of 98\% tetrafluoroethane and 2\% sulfur hexafluoride and are operated in avalanche mode at a continuous flow of 2 l/h and atmospheric pressure. A new bunch of 27 MRPCs have been produced in 2017 for the observatory upgrade: they are again 6 gaps MRPCs with a thinner gap size of 250 $\mu$m \citep{proc_ma}.\\
Recent EU regulations set an upper limit for the \textit{"Global Warming Potential (GWP)"} allowed in gas-operated devices \cite{eu_reg}. The mixture presently adopted in the EEE telescopes has a GWP around 1990, demanding for the search of an alternative.
In order to provide a gas mixture satisfying the EU restrictions, several gas mixtures have been explored, that will be presented in the following sections.
These tests represent the first study in this direction performed on MRPCs operated at low rate, as the one typical of cosmic muons detection with these devices (100 Hz/m$^2$).

\section{Tests performed}
\label{sec::tests}
\begin{figure}[t]
\centering 
\includegraphics[width=0.7\textwidth]{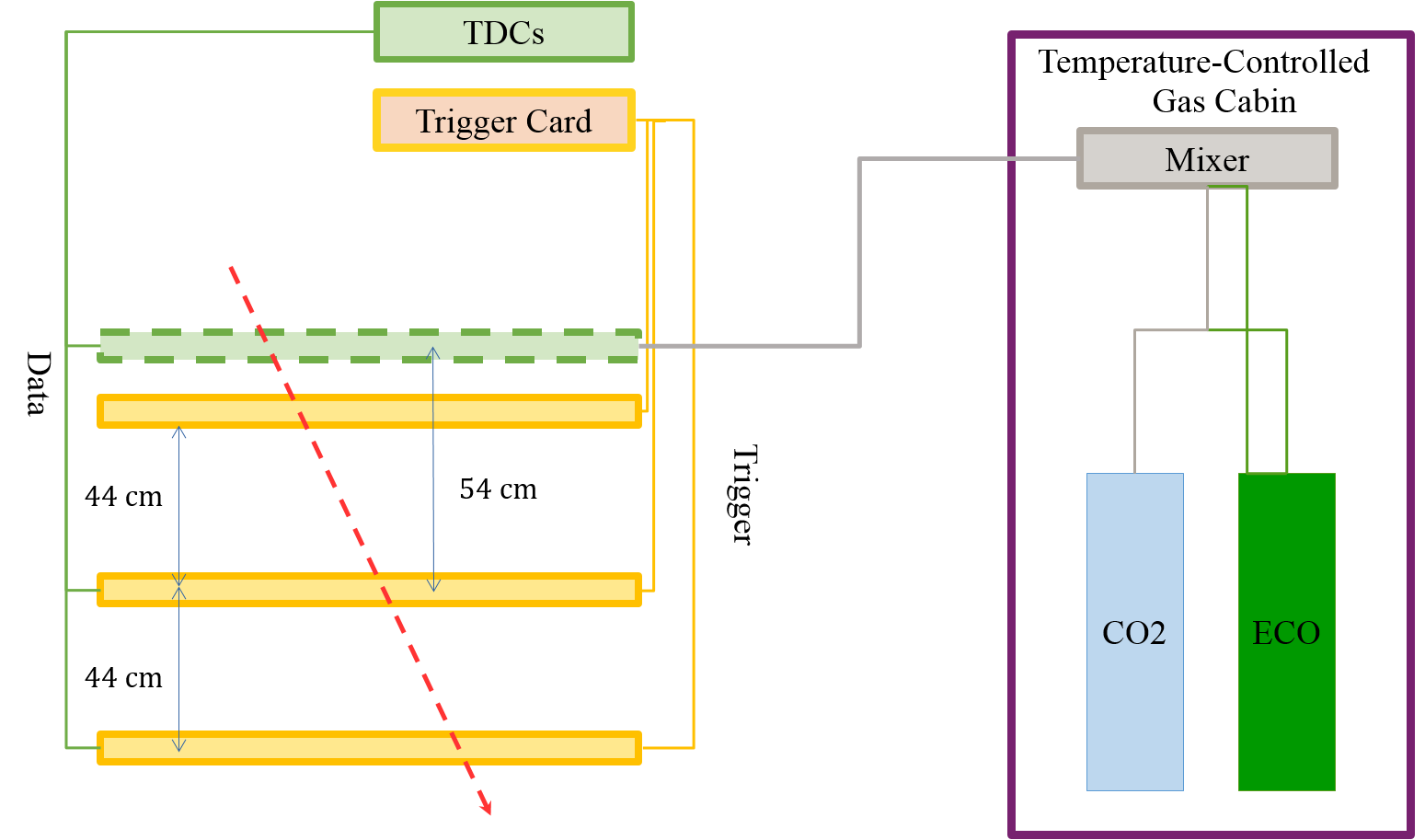}
\caption{\label{fig::setup} Experimental setup.}
\end{figure}
The tests have been performed by detecting and tracking cosmic muons with one of the telescopes of the EEE network installed at CERN (\texttt{CERN-01}).
The setup used is shown in Fig. \ref{fig::setup}: the three chambers of \texttt{CERN-01} - represented in yellow - are operated with the nominal mixture and provide the trigger for the event acquisition. Good events are represented by muons crossing its three chambers.
Events readout is then performed on the two bottom-most chambers of \texttt{CERN-01} and on the chamber under test, positioned on the top of the telescope and represented in green. The latter is operated with the mixture under test, provided by an external mixer.
%
Different combinations have been explored, as reported in Tab. \ref{tab::mixtures}.
\begin{table}[b!]
	\center
	\begin{tabular}{|c|}
		\hline
		Tested Mixtures\\
		\hline \hline
		Pure R1234ze \\
		R1234ze~(90\%) + CO$_{2}$~(10\%) \\
		R1234ze~(80\%) + CO$_{2}$~(20\%) \\
		R1234ze~(50\%) + CO$_{2}$~(50\%) \\
    	R1234ze~(95\%) + SF$_{6}$~(5\%) \\
		R1234ze~(98\%) + SF$_{6}$~(2\%) \\
		R1234ze~(99\%) + SF$_{6}$~(1\%) \\
		CO$_{2}$~(100\%) \\
		CO$_{2}$~(98\%) + SF$_{6}$~(2\%) \\
		\hline
	\end{tabular}
	\caption{Tested gas mixtures.}
	\label{tab::mixtures}
	\endcenter
\end{table}
%
\subsection{Pure R1234ze}
\label{ssec::pure_eco}
\begin{figure}[htbp]
\centering
\includegraphics[width=0.75\textwidth]{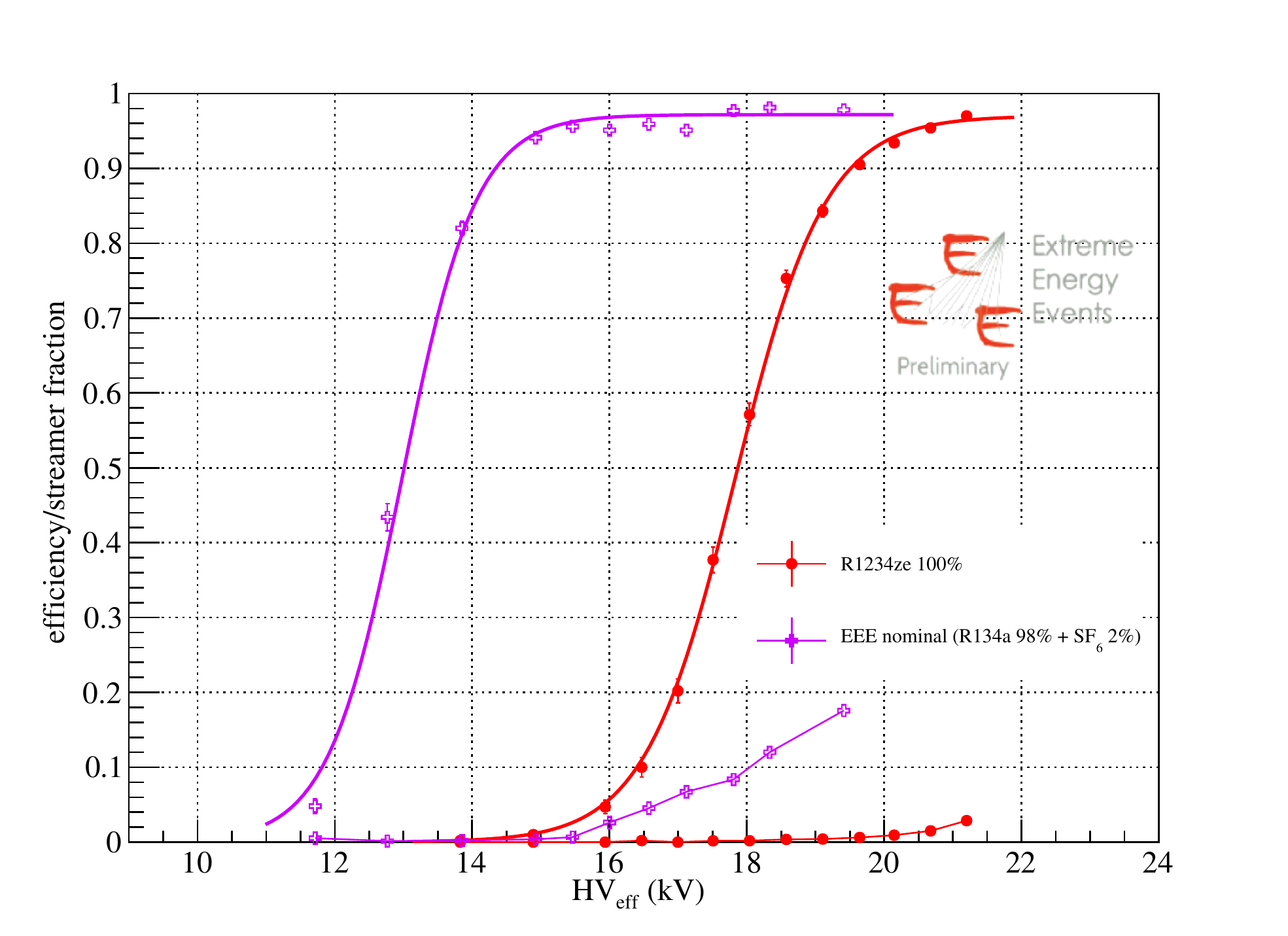}
\caption{\label{fig::eco} Efficiencies and streamer fraction for the nominal EEE gas mixture and one made of R1234ze.}
\end{figure}

The dominant component of the EEE nominal mixture is the
tetrafluoroethane (C$_2$H$_2$F$_4$), with a GWP=1430. 
In recent years, the tetrafluoropropene (C$_3$H$_2$F$_4$, coming in two allotropic forms, generally labeled as R1234yf, and R1234ze, of which R1234ze has been used, since the other is slightly flammable), has emerged as a good substitute for the former, thanks to its sensibly lower GWP=6.
The first mixture analyzed was composed of 100\% R1234ze.
In Fig. \ref{fig::eco}, the efficiency and the streamer fraction obtained with the $R1234ze$ are shown, in comparison with the nominal ones. The streamer fraction has been evaluated as the ratio between the number of hits with a high cluster size ($\geq$ 5) and the total number of hits.
The new gas shows excellent quenching properties (streamer component stays below 5\% in the whole range) and exhibits a stable efficiency plateau; however, the latter is reached at a $HV_{eff}\footnote{$HV_{eff}= HV\times\frac{p_0}{p}\frac{T}{T_0}$, with $p_0$=1000 mbar and $T_0$=298.15 K representing standard temperature and pressure, and $p$ and $T$ being the ones in the detector room.}\approx21$kV, at the limit of the operability regime for the EEE MRPCs.
%
%
%
\subsection{Mixtures of CO$_2$ and R1234ze}
\label{ssec::eco_co2}
\begin{figure}[htbp]
\centering
\includegraphics[width=0.75\textwidth]{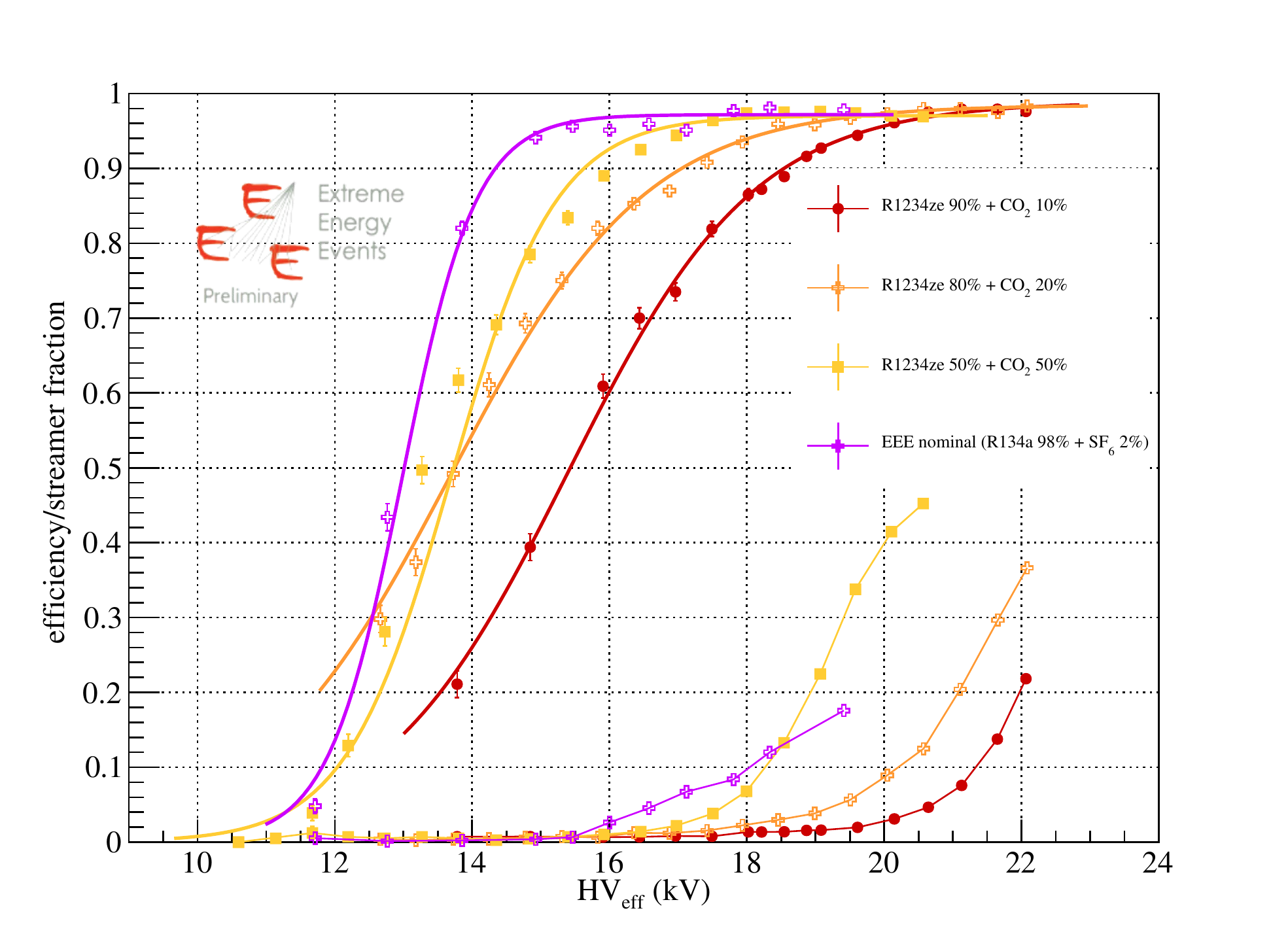}
\caption{\label{fig::eco_co2} Efficiencies and streamer percentage for the different mixtures of CO$_2$ and R1234ze tested.}
\end{figure}

A natural choice to lower the operating high-voltage of tetrafluoropropane gas mixtures is to add CO$_2$. Being the GWP of CO$_2$ the reference value to define the global warming potential of other gases, \textit{i.e.} GWP$_{CO_2}$=1, no limitations in the addition of this gas derive from ecological regulations.
As a consequence, different $CO_2$ percentages have been tested, ranging from a minimum of 10\% to a maximum of 50\%, in combination with $R1234ze$.
Fig. \ref{fig::eco_co2} shows the results obtained with the different combinations.
While the high voltage needed to reach the plateau is sensibly lowered with respect to the pure $R1234ze$ case and a stable working region can be identified, the presence of $CO_2$ produces a noisy configuration, with the streamer component getting too high when entering the efficiency plateau. However, a possibile working point at $17-18$ kV for the $R1234ze(50\%)$ + $CO_{2}$(50\%) mixture can be identified, despite the relatively high streamer percentage.
%
\subsection{Mixtures of R1234ze and SF$_6$}
\label{ssec::eco_sf6}
\begin{figure}[h]
\centering
\includegraphics[width=0.75\textwidth]{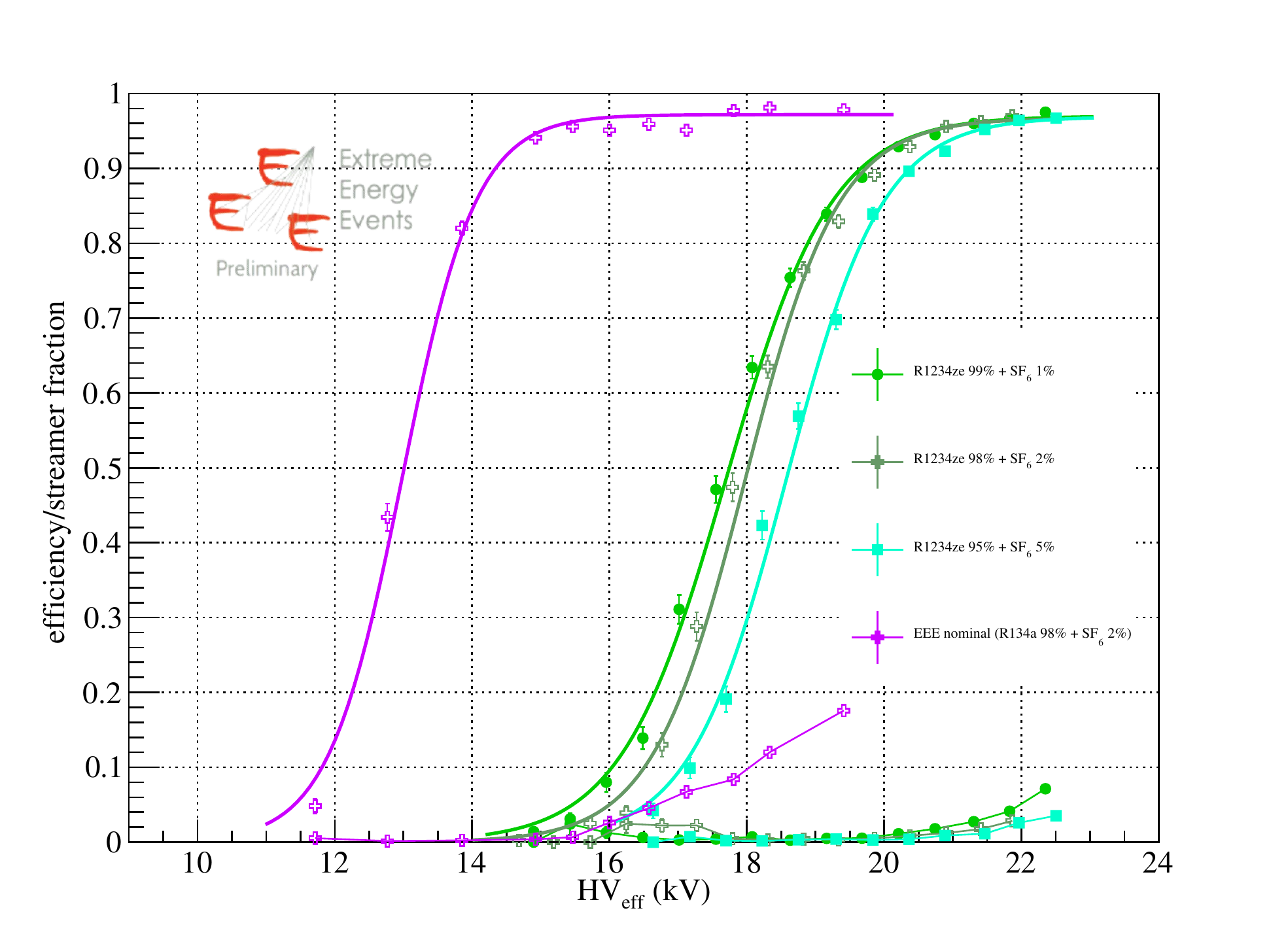}
\caption{\label{fig::sf6} Efficiency and streamer fraction for the R1234ze and SF$_6$ mixtures tested.}
\end{figure}

In order to minimize the streamer component, the addition of a gas with good quenching properties is needed. Toward this direction, alternative mixtures have been tested, composed of R1234ze and SF$_6$.
However, due to the high global warming potential of the $SF_6$ ($GWP_{SF_6}$=23900), its amount has to be kept very limited.
Fig. \ref{fig::sf6} shows the results obtained exploring different compositions. On the one hand, the presence of $SF_6$ confirms its quenching capability, with a streamer component at the level of a few percent; however, it is demanding in terms of operating voltage, since the efficiency plateau is reached only for very high $HV_{eff}$ values, at the limit of the allowed operating conditions for the EEE MRPCs.
The data trend allows, however, to identify a possible strategy: since the SF$_6$ is very effective as a quencher already at very low percentage, and since the diminishing of the amount of SF$_6$ shifts toward lower operating voltages the efficiency plateau, a further combination can be explored, \textit{i.e.} a mixture R1234ze (99.5\%) + SF$_{6}$ (0.5\%), that would fulfill EU regulations and with a working region where the $HV_{eff}$ is not too high. This will be done in the near future.
%
\subsection{Mixtures of CO$_2$ and SF$_6$}
\label{ssec::co2_based}
\begin{figure}[h]
\centering
\includegraphics[width=0.75\textwidth]{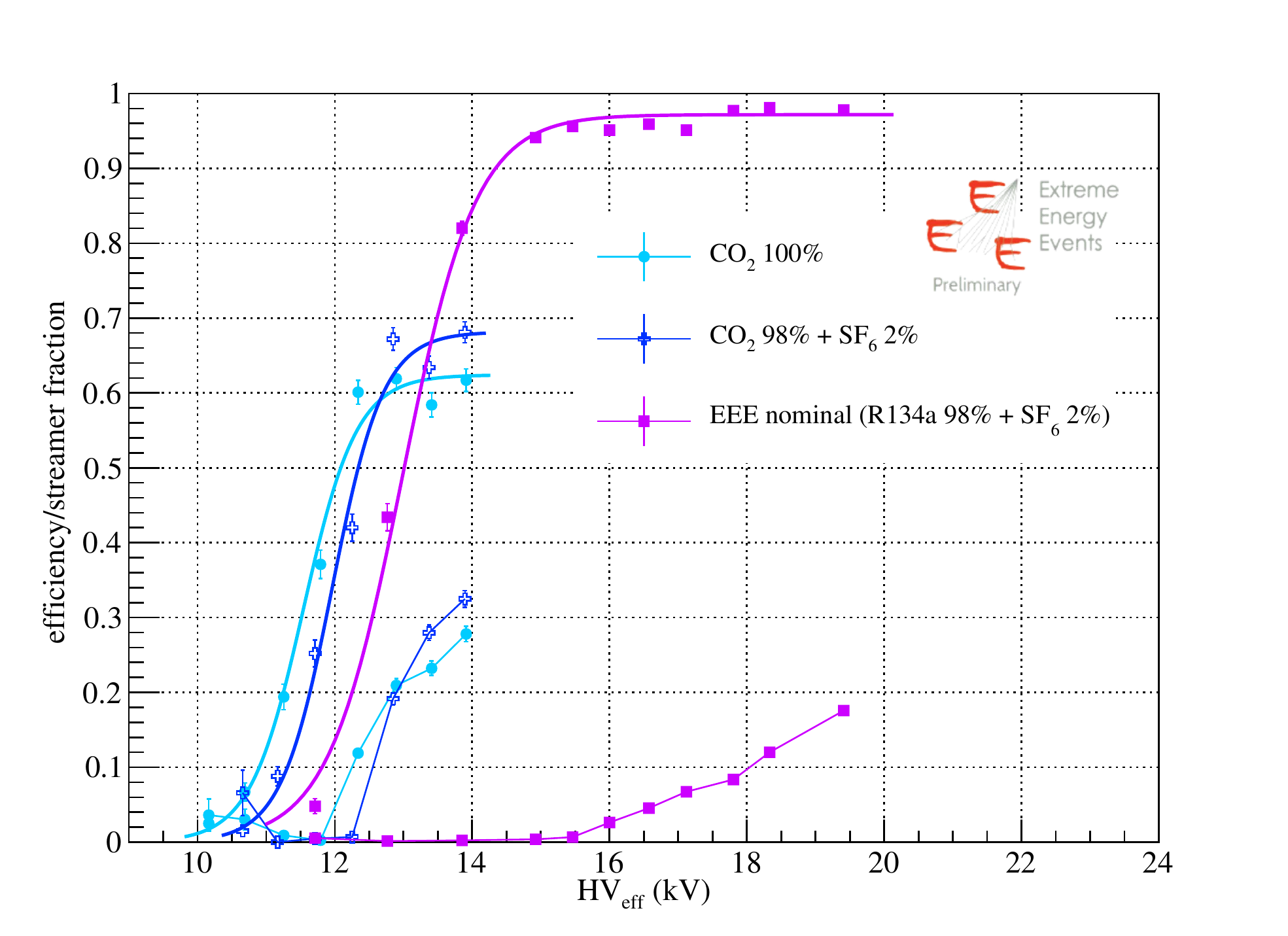}
\caption{\label{fig::co2_based} Efficiency and streamer probability for the CO$_2$ and SF$_6$ mixtures tested.}
\end{figure}

Combining the information collected with the previous studies, a further combination has been tested, with mixtures based on CO$_2$ and SF$_6$.
In principle, the excellent quenching properties of SF$_6$ and the low operating voltage allowed by CO$_2$ could provide an optimal configuration, with both $HV_{eff}$ and streamer percentage kept very low. The results are shown in Fig. \ref{fig::co2_based}: while the plateau is reached at a very low voltages, the streamer component is diverging. Furthermore, the efficiency turns out to be too low, so this combination has to be discarded.
%
%
\begin{figure}[h!]
\centering
\includegraphics[width=0.75\textwidth]{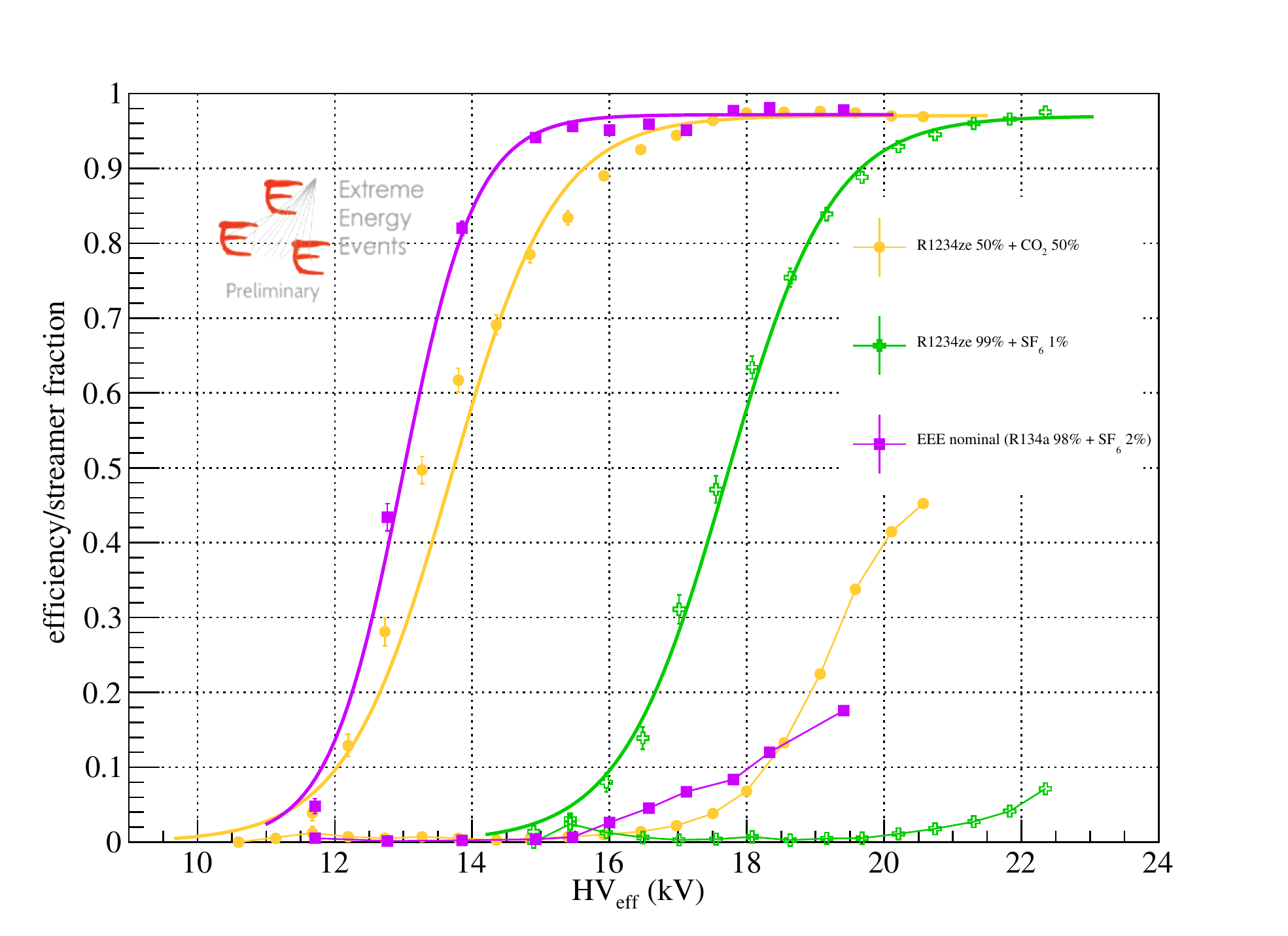}
\caption{\label{fig::best} Most promising mixtures tested: efficiencies and streamer percentage.}
\end{figure}
\newpage
\section{Conclusions}
Fig. \ref{fig::best} summarizes the efficiencies and streamer components of the most promising mixtures, \textit{i.e.} $R1234ze(50\%)$ + $CO_{2}$(50\%) and $R1234ze(99\%)$ + $SF_{6}$(1\%). In the near future, some of the EEE stations will be equipped with these mixtures, and telescope performances and stability will be investigated on a long-term perspective.
In addition to it, further tests have been planned, that will explore alternative gases and mixtures such as $CF3I$, $R1234ze(99,5\%) + SF_{6}$(0,5\%), $R1234ze + He$.


\begin{thebibliography}{99}
%
\bibitem{eee_project}
\small \emph{Progetto "La Scienza nelle Scuole" - EEE: Extreme Energy Events, Societ\`a Italiana di Fisica (Bologna)}, \emph{A. Zichichi} 3 May 2004; 2nd edition, 1 April 2005; 3rd edition, 20 April 2012
%
\bibitem{CFsite}
\emph{http://www.centrofermi.it/EEE}
%
\bibitem{coinc1}
\emph{M. Abbrescia et al. (EEE Collaboration)}, \emph{First detection of extensive air showers with the EEE experiment}, {\bf Nuovo Cim. B125} (2010) 243
%
\bibitem{coinc2}
\emph{M. Abbrescia et al. (EEE Collaboration)}, \emph{Search for long distance correlations between extensive air showers detected by the EEE network}, {\bf Eur. Phys. J. Plus} (2018) 133: 34
%
\bibitem{forbushDec2}
\emph{M. Abbrescia et al. (EEE Collaboration)}, \emph{Observation of the February 2011 Forbush decrease by the EEE telescopes }, {\bf Eur. Phys. J. Plus} (2011) 126: 61
%
\bibitem{forbushDec3}
\emph{M. Abbrescia et al. (EEE Collaboration)}, \emph{The EEE experiment project: status and first physics results }, {\bf Eur. Phys. J. Plus} (2013) 128: 62
%
\bibitem{anisotropies}
\emph{M. Abbrescia et al. (EEE Collaboration)},
\emph{Looking at the sub-TeV sky with cosmic muons detected in the EEE MRPC telescopes}, {\bf Eur.Phys.J.Plus 130} (2015) 187
%
\bibitem{upward}
\emph{M. Abbrescia et al. (EEE Collaboration)}, \emph{A study of upward going particles with the Extreme Energy Events telescopes }, {\bf Nucl. Instr. and Meth. A 816} (2016) 142-148
%
\bibitem{proc_ma}
\emph{M. Abbrescia}, \emph{"The upgrade of the Extreme Energy Events experiment"}, these proceedings.
%
\bibitem{proc_ddg}
\emph{D. De Gruttola}, \emph{"Performance of the Multigap Resistive Plate Chambers of the Extreme Energy Events Project"}, these proceedings.
%
%
\bibitem{eee_perf_paper}
\emph{M. Abbrescia et al. (EEE Collaboration)}, \emph{The Extreme Energy Events experiment: an overview of the telescopes performance} {\bf https://arxiv.org/pdf/1805.04177.pdf}
%
\bibitem{eu_reg}
Regulation (EU) No 517/2014 of the European Parliament and of the Council on fluorinated greenhouse gases and repealing Regulation (EC) No 842/2006.
%
%
\end{thebibliography}
\end{document}